\documentclass[aps,prl,twocolumn,groupedaddress,amsmath,amssymb]{revtex4-1} 
\usepackage{graphicx}  	
\usepackage{dcolumn}  	
\usepackage{bm}        	
\usepackage{verbatim}   	
\usepackage{color}

\begin{document}

\title{Grazing incidence interferometry for rough convex aspherics}

\author{Johannes Schwider}
\author{Christine Kellermann}
\author{Norbert Lindlein}
\author{Sergej Rothau}
\affiliation{ 	 
   Institute of Optics, Information, and Photonics,\\
   Friedrich-Alexander-University Erlangen-N\"urnberg (FAU),\\
   Staudtstr. 7/B2, 91058 Erlangen, Germany\\
   }
\author{Klaus Mantel}  
\affiliation{   
   Max Planck Institute for the Science of Light,\\
   Staudtstr 2, 91058 Erlangen, Germany\\
   }
   
\date{\today}
\begin{abstract}
Grazing incidence interferometry has been applied to rough planar and cylindrical surfaces. As suitable beam splitters diffractive optical phase elements are quite common because these allow for the same test sensitivity for all surface points. But a rotational-symmetric convex aspheric has two curvatures which reduces the measurable region to a meridian through the vortex of the aspheric, which is in contrast to cylindrical surfaces having a one-dimensional curvature which allows the test of the whole surface in gracing incidence. The meridional limitation for rotational-symmetric aspherics nevertheless offers the possibility to measure single meridians in a one-step measurement. An extension to the complete surface can be obtained by rotating the aspheric around its vortex within the frame of the test interferometer.
\end{abstract}

\maketitle

\section{Introduction}

\begin{figure}[htbp]
\centering
\fbox{\includegraphics[width=0.92\linewidth]{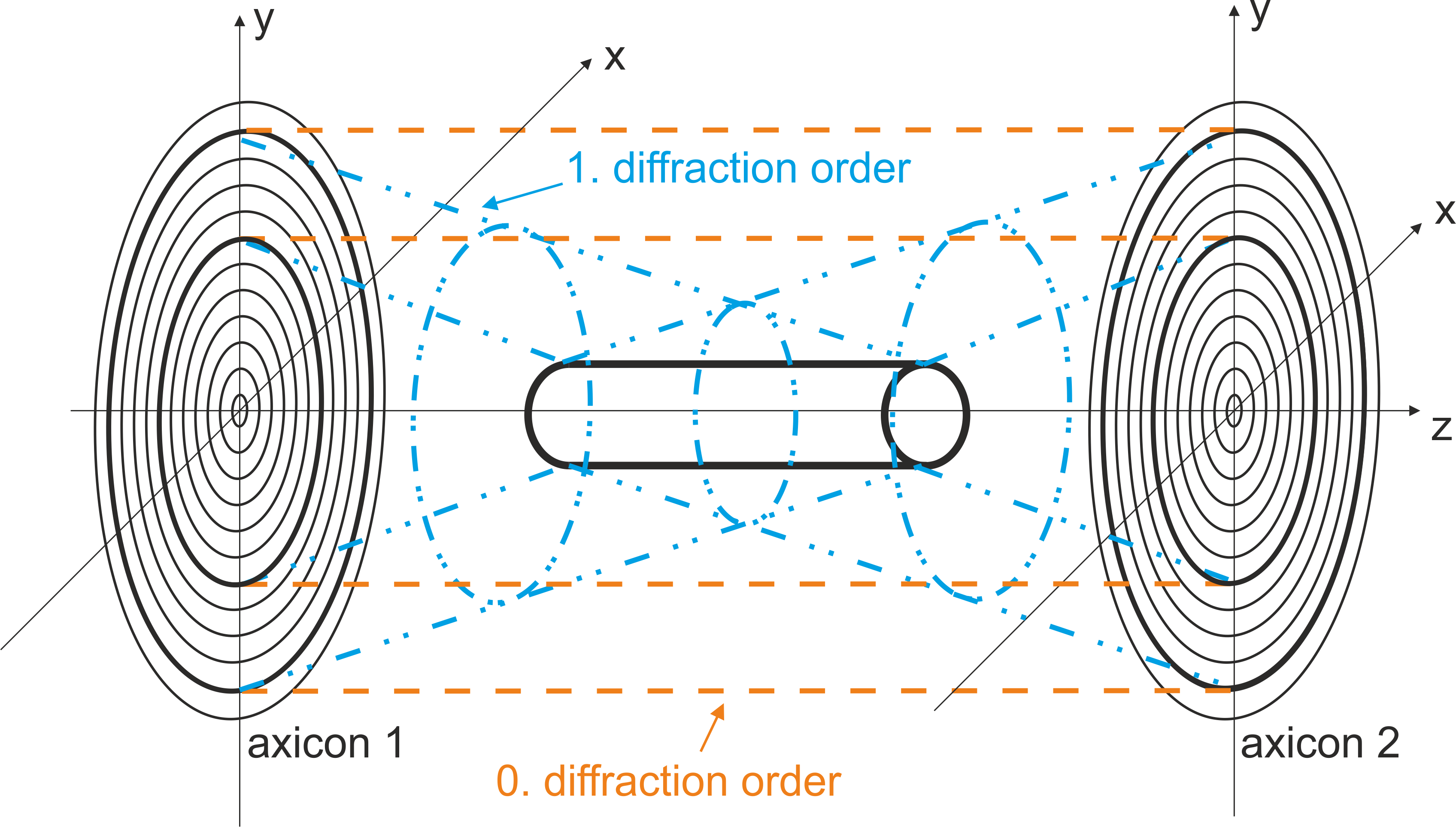}}
\caption{Core of a grazing incidence interferometer for the testing of a cylindrical specimen. The null lenses are realized as diffractive axicons. As a reference wave, the undiffracted light is used. The balance between the undiffracted and diffracted waves can be steered through the depth of the phase structure.}
\label{fig:FIG1a}
\end{figure}

Interferometric tests are the most accurate means for the control of the production process of aspheric surfaces in optics. This is true for smooth surfaces in the first place. However, optical surfaces resulting from the grinding step cannot so easily be tested with interferometric methods. Only by increasing the test wavelength into the IR-region \cite{Kwon} or through the use of gracing incidence \cite{Abramson} of the probing wave onto the surface to be tested one could enable interferometric measurements. Here, the application of grazing incidence interferometry to rotationally symmetric aspherics shall be discussed as a generalization of the test of developable surfaces like cylindrical surfaces \cite{Dresel1,Schwider} (see Fig. \ref{fig:FIG1a}). 

The discussion will deal with the limitations and the implications due to gracing incidence if curved surfaces with positive Gaussian curvature, like rough convex aspherics, shall be measured. 

\begin{figure}[htbp]
\centering
\fbox{\includegraphics[width=0.92\linewidth]{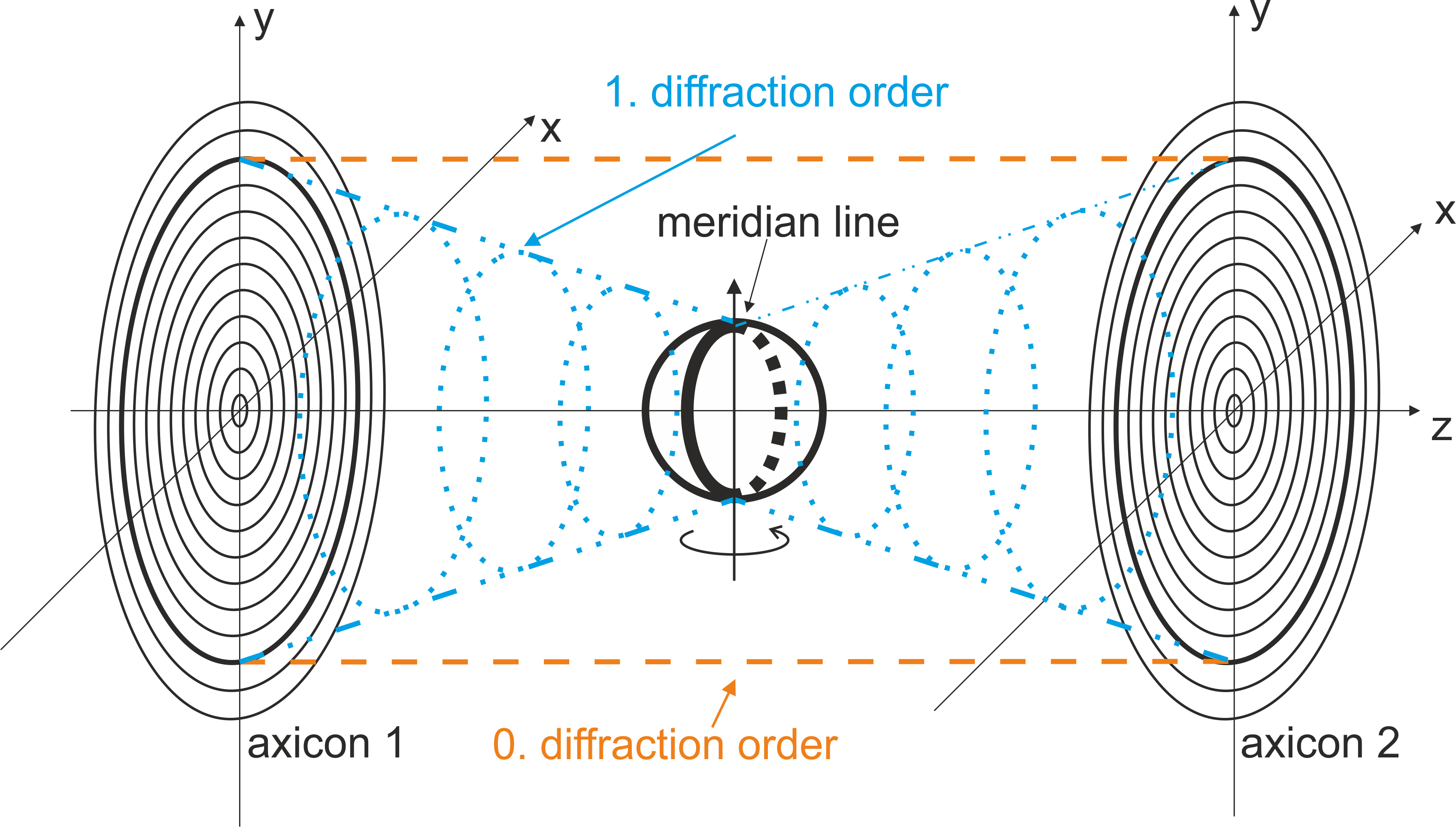}}
\caption{Sketch of the core of a grazing incidence interferometer using two diffractive axicons as local null lenses for the test of a spherical ball lens.}
\label{fig:FIG1b}
\end{figure}

Birch \cite{Birch} has applied the grazing incidence test after Abramson \cite{Abramson} to rather shallow aspherics in the close surrounding of the vertex. Such a test is of course of only limited value for steeper aspherics because of geometrical reasons concerning sensitivity variations and obscuration of light rays.  
Since convex cylindrical surfaces can be tested as a whole in gracing incidence \cite{Brinkmann} it is evident that the same test geometry can be applied to a general convex rotational symmetric surface for a meridian through the vertex of the surface (see Fig. \ref{fig:FIG1b}).

Measuring aspherics in grazing incidence offers several advantages over commonly used tests in normal incidence. In the latter case, the diameter of the diffractive optical element (DOE) used as null lens gets exceedingly large for high aperture aspherics, while the period of the structures at the edge of the DOE may become extremely small, making the DOE fabrication difficult or even impossible. In grazing incidence, on the other hand, the DOE period is constant and, depending on the desired sensitivity, generally in the range of $5\mu \textrm{m}$ to $20\mu \textrm{m}$, posing no problems for the fabrication process. Therefore, in grazing incidence, the full aperture even of high aperture aspherics can be measured. Owing to the reduced effective wavelength, which is equal to the DOE period, the surface under test may be rough and can be measured before fine polishing. These advantages, however, come with a price: for rotationally symmetric specimens, the procedure works only locally, in a small area around the meridian of the aspheric. For a full measurement of the surface, the aspheric has to be rotated while a series of measurements is taken, which have to be merged afterwards.

\section{Principle of the proposed grazing incidence test}

To apply grazing incidence interferometry to the measurement of rough, high aperture aspherics, we propose that the probe wave impinging onto the surface is produced through diffraction at a DOE having equidistant parallel curve structures to the meridional curve of the surface under test. The interference pattern is then confined to a small region of the same shape. The following imaging telescope images the vertex region of the aspheric onto a CMOS camera chip. The interference pattern can be evaluated by phase shifting techniques. The necessary phase shift can be introduced by a suitable axial movement of one of the DOE's delivering a set of intensity patterns.

\begin{figure}[htbp]
\centering
\fbox{\includegraphics[width=0.92\linewidth]{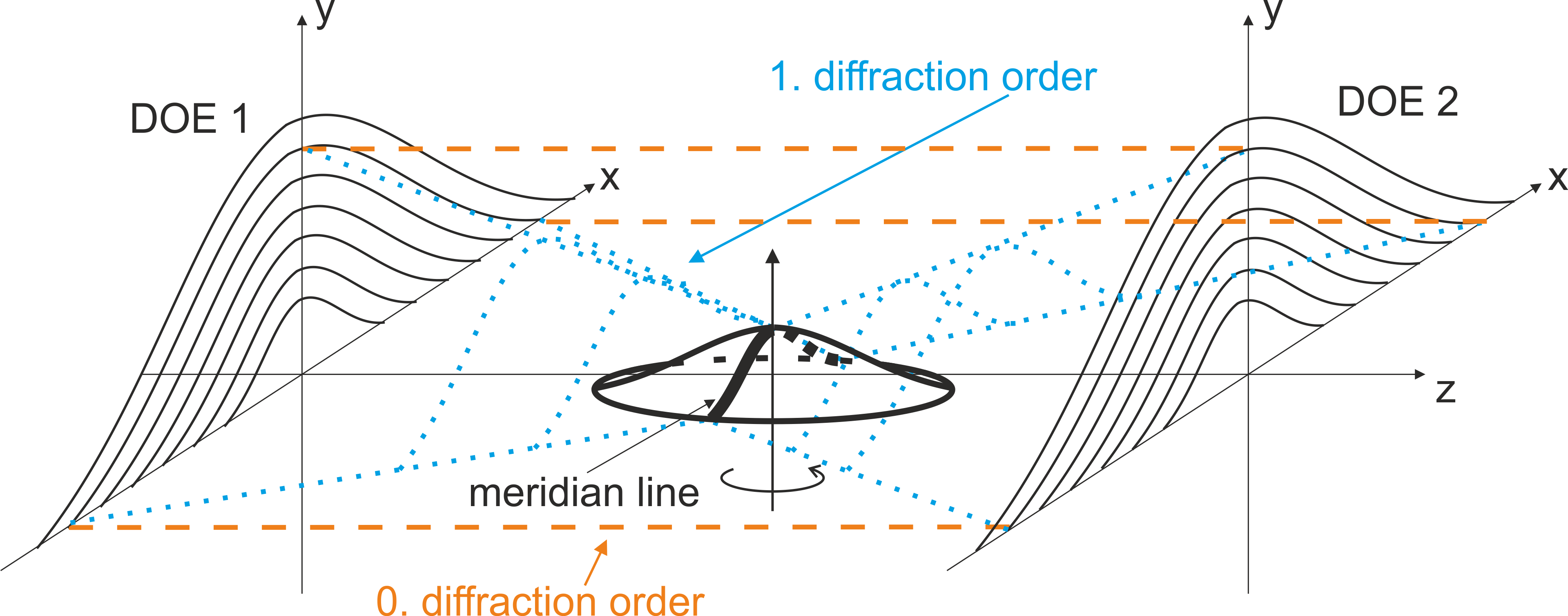}}
\caption{Core of a grazing incidence interferometer using two DOE as local null lenses with an aspheric lens as a specimen.}
\label{fig:FIG1c}
\end{figure}

\section{Simulations for a spherical ball lens}

For the discussion of the basic features of the grazing incidence test proposed here, it is sufficient to restrict simulations to spherical surfaces because the mathematics behind the principle is very similar. For a first simulation, a spherical ball lens with a diameter of 50mm has been chosen, a dimension within the typical margin of aspherics being part of optical imaging systems. Such a ball lens represents the most extreme geometry imaginable from the viewpoint of the numerical aperture of the lens under test.

\begin{figure}[htbp]
\centering
\fbox{\includegraphics[width=0.92\linewidth]{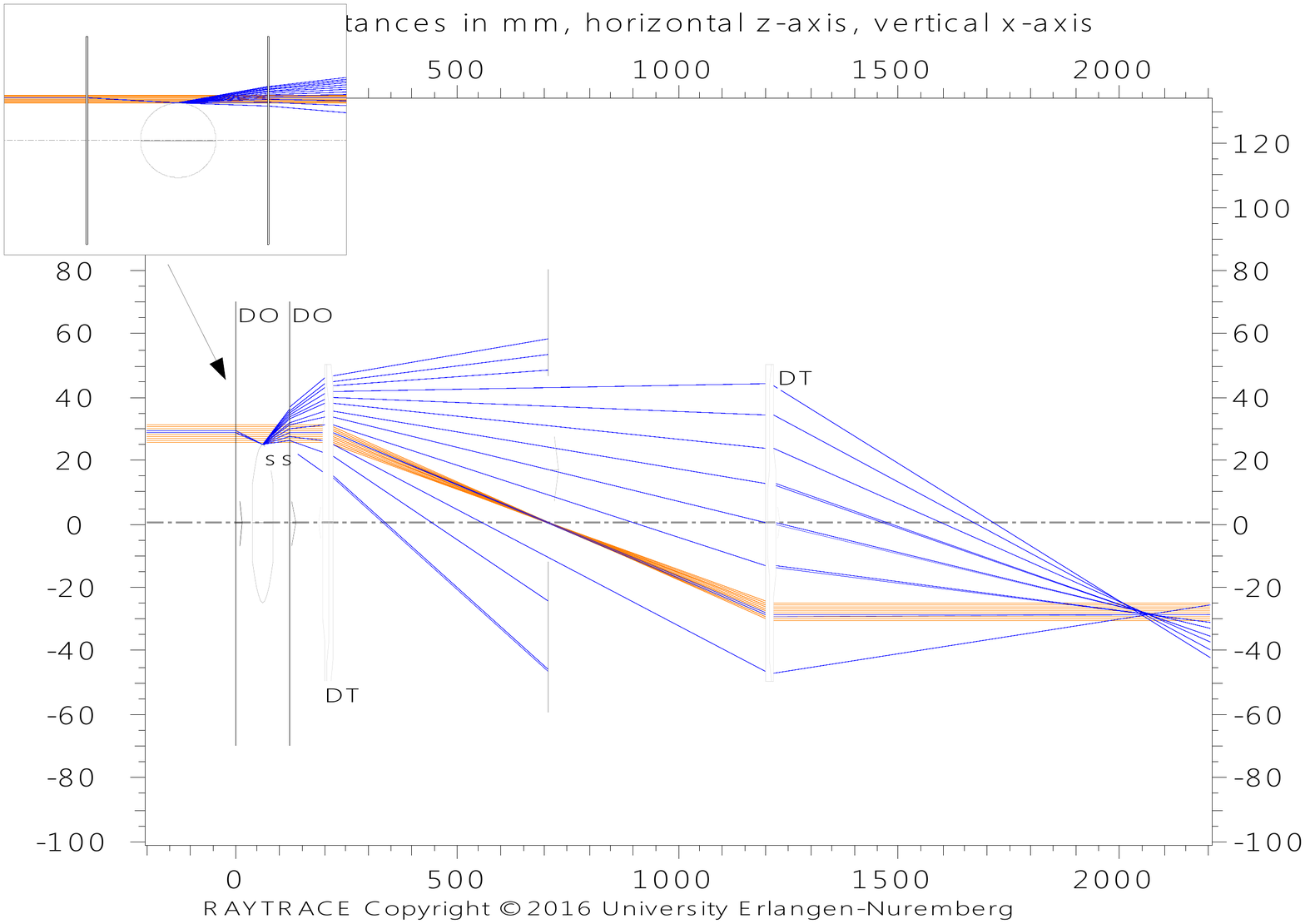}}
\caption{Overview of the simulated system with both axicons, the spherical reflective surface under test, and the imaging telescope.}
\label{fig:FIG2a}
\end{figure}

A ray trace simulation shall give an insight into the wave front propagation in the interferometer and, via the imaging telescope, to the chip plane of the camera.
Figure \ref{fig:FIG2a} shows an overview of the whole simulated system, including a subplot of magnified region. 

\begin{figure}[htbp]
\centering
\fbox{\includegraphics[width=0.92\linewidth]{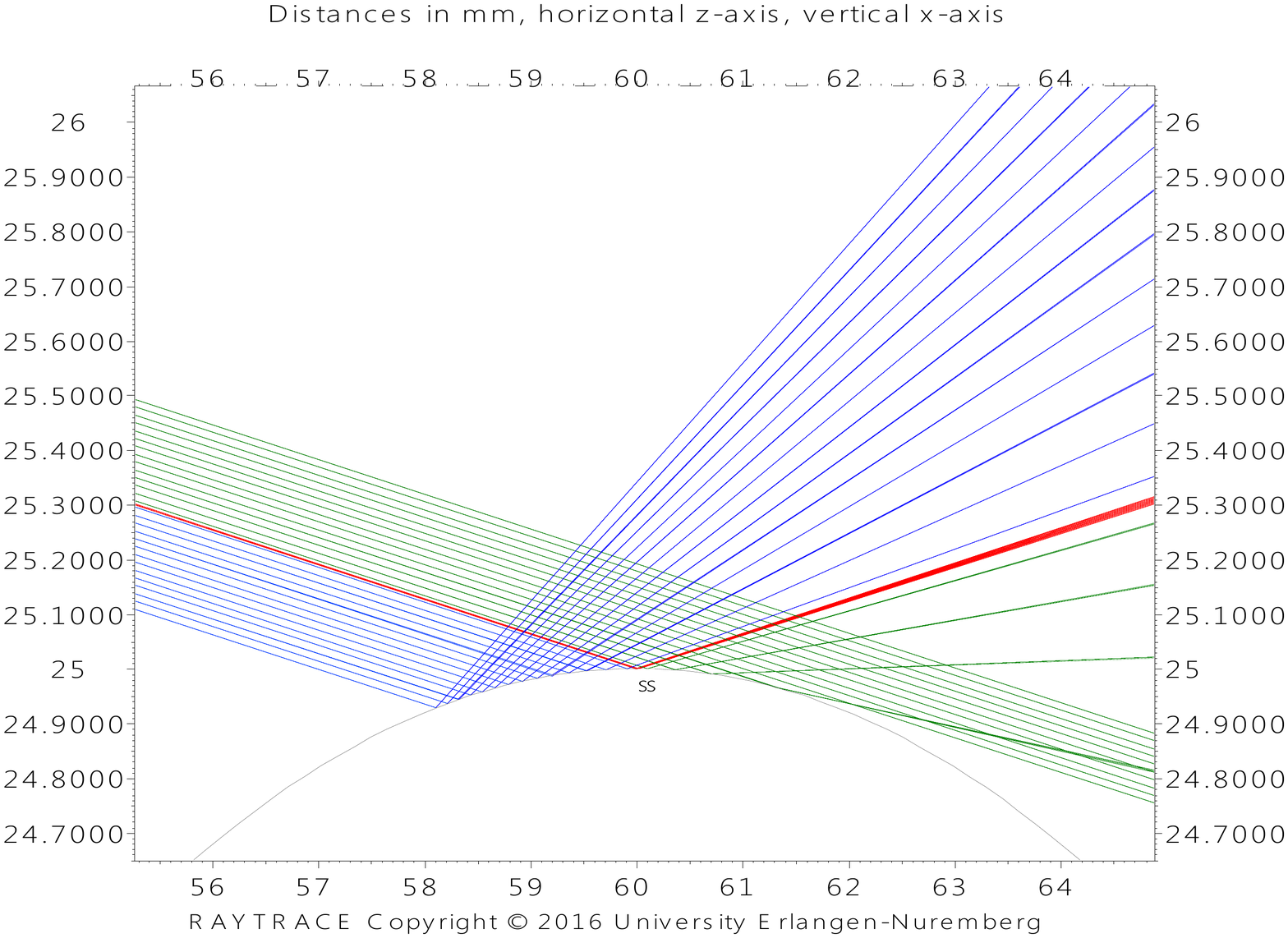}}
\caption{Magnified meridional area in the vicinity of the vertex of the sphere.}
\label{fig:FIG2b}
\end{figure}

Figure \ref{fig:FIG2b} shows the meridional region. The area that can be measured is located in an asymmetric fashion around the meridional curve. Furthermore, the light rays are distorted in a nonlinear way, leading to a nonlinear relationship between the coordinate systems of the DOE and the surface.

\begin{figure}[htbp]
\centering
\fbox{\includegraphics[width=0.92\linewidth]{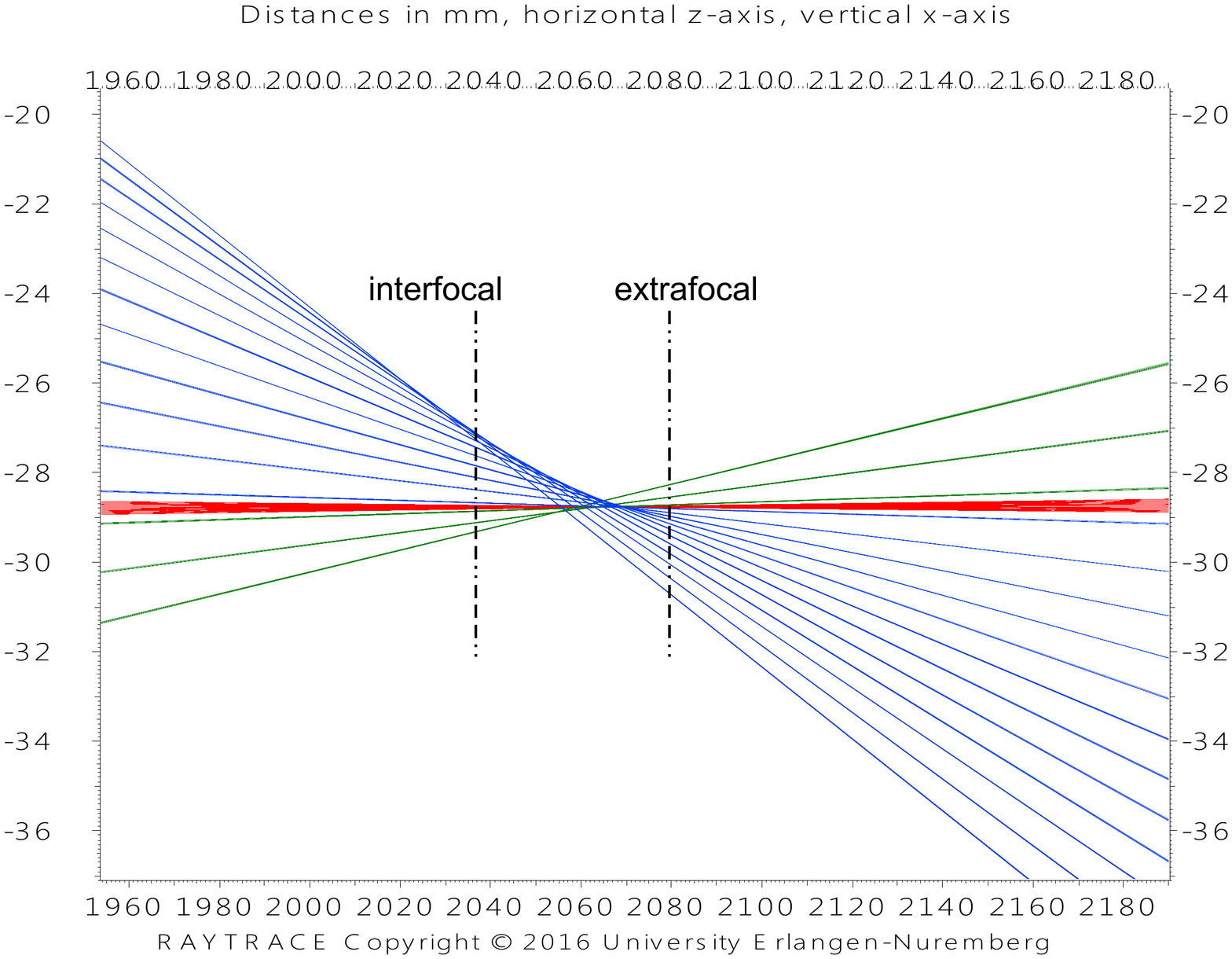}}
\caption{Ray bundle in the vicinity of the detector area.}
\label{fig:FIG2c}
\end{figure}

\begin{figure}[htbp]
\centering
\fbox{\includegraphics[width=0.92\linewidth]{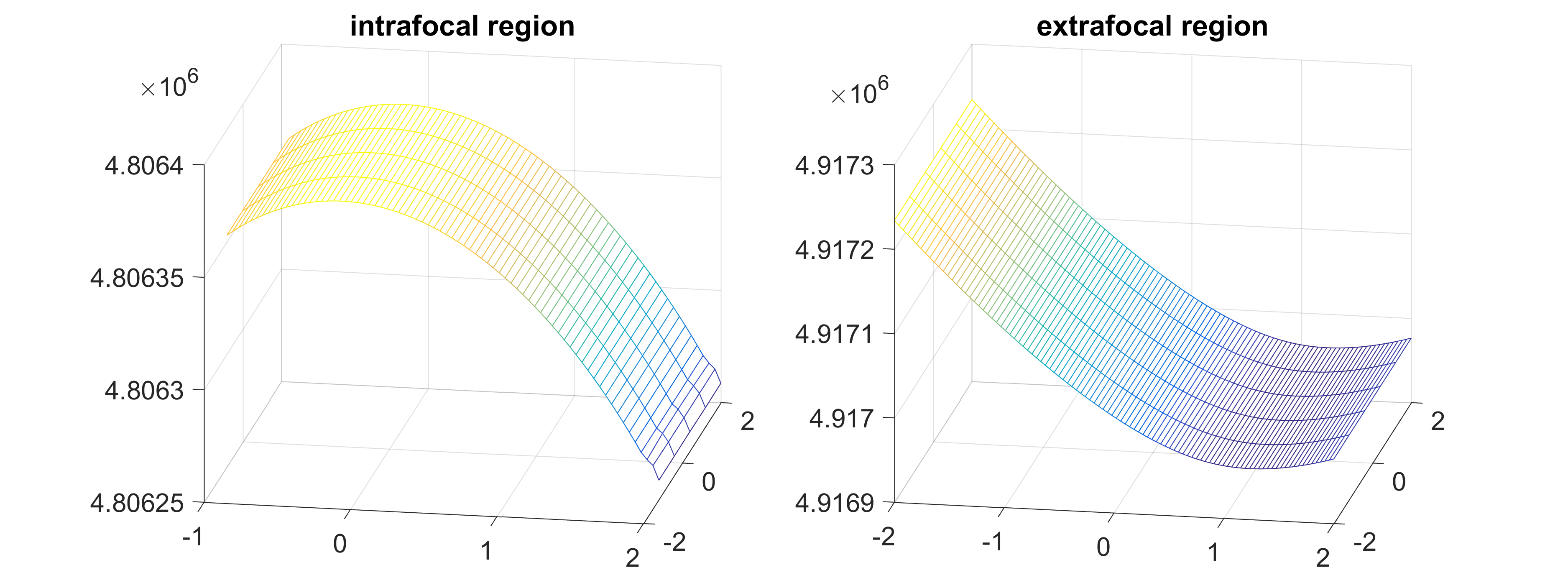}}
\caption{Shape of the wave front in the intrerfocal (\textit{left}) and extrafocal (\textit{right}) region of the imaging telescope.}
\label{fig:FIG3}
\end{figure}

The ray tracing simulations of Fig. \ref{fig:FIG2c} confirm the presence of a highly distorted wave front in the image plane. For a small region around the vertex, intra- and extrafocal polynomial representations can be derived through lsq-fits of the optical path length data, shown in Fig. \ref{fig:FIG3}. 


\begin{figure}[htbp]
\centering
\fbox{\includegraphics[width=0.55\linewidth]{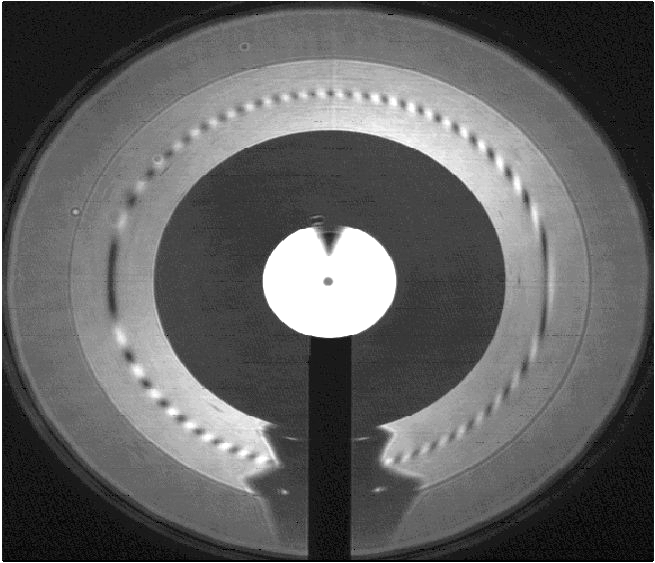}}
\caption{Interference pattern of a cylindrical surface under spatially partial coherent illumination. Due to the inversion of the probe wave front through reflection at the cylinder surface the fringe region is restricted to the plane of symmetry between the DOE´s of the setup (see Fig. \ref{fig:FIG1a}).} \label{fig:FIG4}
\end{figure}

One very essential step in the reconstruction of the wave front to be measured and, therefore, also for the measurement of surface features, is the unwrapping step which requires a two-dimensional region for unequivocal reconstructions. Figure \ref{fig:FIG4} gives an example for the fringe patterns usable for real measurements, where a grazing incidence interferogram of a cylindrical rod is shown, taken with partially coherent light produced with the help of a rotating scatterer in the laser path. The inversion of the probe wave through reflection at the meridional plane in relation to the reference wave leads to the localisation of the fringes to this plane. There is an obvious gain in the smoothness of the fringes which will alleviate the unwrapping process in the final measurements. A great deal of knowledge gathered in our previous work on the elimination of misalignment aberrations for rod objects \cite{Dresel2} can here be applied. 

\subsection{Future prospects}

The extension of the meridional test onto the whole surface seems possible by measuring different meridians one after the other. For matching the data, the fact can be used that measurements taken in two azimuths being $180^\circ$ apart show the same deviation picture but mirrored at the vertex of the surface under test. Such additional data should alleviate the reduction of ambiguities due to the successive data sampling process for the whole surface. Such ambiguities might be caused through small misadjustments of the aspheric relative to the axis of the rotating stage. Erroneous data resulting from imperfect adjustment can be detected on the one hand and on the other corrected via software.

\end{document}